\documentstyle[12pt]{article}
\begin{document}
\def\singlespacing{\baselineskip=14pt}
\def\doublespacing{\baselineskip=18pt}
\doublespacing

\pagestyle{empty}

\noindent cond-mat/9603092 \hfill Submitted to Physical Review E

\bigskip

\begin{center}
\begin{Large}
{\bf Kinetics of fragmentation-annihilation processes} \\
\end{Large}
\bigskip
\bigskip
J.\ A.\ N.\ Filipe\footnote{email: Joao.Filipe@brunel.ac.uk. Address from 
April 1, 1996: Biomathematics \& Statistics Scotland, The University 
of Edinburgh, James Clerk Maxwell Building, The King's Buildings, Edinburgh 
EH9 3JZ, UK.} and G.\ J.\ Rodgers\footnote{email: G.J.Rodgers@brunel.ac.uk} \\
\bigskip
Department of Physics, Brunel University, Uxbridge, Middlesex UB8 3PH, UK \\
\bigskip
\bigskip
\bigskip

{\bf Abstract}
\end{center}

\doublespacing

We investigate the kinetics of systems in which particles of one species
undergo binary fragmentation and pair annihilation. In the latter,
non-linear process, fragments react at collision to produce an inert
species, causing loss of mass.
We analyse these systems in the reaction-limited regime by solving a
continuous model within the mean-field approximation.
The rate of fragmentation, for a particle of mass $x$ to break into
fragments of masses $y$ and $x-y$, has the form $x^{\lambda-1}$
($\lambda>0$), and the annihilation rate is constant and independent
of the masses of the reactants.
We find that the asymptotic regime is characterized by the annihilation
of small-mass clusters, with the cluster-density decaying as in
pure-annihilation and the average cluster-mass as in pure-fragmentation.
The results are compared with those for a model with linear mass-loss
(i.e.\ with a sink rather than a reaction).

We also study more complex models, in which the processes of fragmentation
and annihilation are controlled by mutually-reacting catalysts.
Both pair- and linear-annihilation are considered.
Depending on the specific model and initial densities of the catalysts, the
time-decay of the cluster-density can now be very unconventional and
even non-universal.
The interplay between the fragmentation and annihilation processes and the
existence of a scaling regime are determined by the asymptotic behaviour
of the average-mass and of the mass-density, which may either decay
indefinitely or tend to a constant value.
We discuss further developments of this class of models and their
potential applications.

\bigskip


\newpage

\pagestyle{plain}
\pagenumbering{arabic}

\section{Introduction}

Irreversible reaction-diffusion processes, such as aggregation or
annihilation of species, occur in many physical, chemical and
biological systems.
In the past decade they have been investigated extensively.
This followed from the realization that spatial fluctuations in the
reactant densities slow down the kinetics and lead to an asymptotic
behaviour at low-dimensions which disagrees with the classical
reaction-rate equations. (The upper critical dimension, $d_c$, is
defined as the spatial dimension below which such disagreement occurs.)
Recently, a combined process of aggregation-annihilation of two
species has been studied exactly in 1 dimension by Krapivsky
\cite{krap93}, and numerically in higher dimensions by Sokolov and
Blumen \cite{sokolov.blumen}.
A class of related models, but with an arbitrary number of species, has
also been investigated at the mean-field level by Ben-Naim and Krapivsky
\cite{bennaim.krap95}. The power-law decay of the cluster-densities has
been understood, and in many cases the cluster-mass distributions have
been found to approach universal scaling forms.

Fragmentation is another irreversible kinetic process with important
applications in nature and technology. These include: polymer degradation,
through shear action \cite{poly.shear}, chemical attack \cite{poly.chem},
or exposure to radiation \cite{poly.rad}; droplet breakup \cite{droplet};
fibre-length reduction \cite{fibre}; fragmentation of colloidal aggregates
\cite{colloids}; and rock crushing and grinding \cite{fibre}.
New insight into the statistical mechanics of fragmentation has followed
 from the work of Ziff and McGrady \cite{ziff.mcgrady85,ziff.mcgrady86},
who solved exactly a class of models with different fragmentation kernels.

The aggregation-fragmentation (or reversible-aggregation) process has been
considered by many authors. This has been applied to studying reversible
polymerization, for instance, by van Dongen and Ernst \cite{vandongen.ernst}
and by Sintes, Toral and Chakrabarti \cite{sintes.etal}.
More generally, mean-field approaches have been used to study the
steady-state scaling properties of reversible-aggregation, for example,
by Family, Meakin and Deutch \cite{family.etal} and by Blackman and
Marshall \cite{blackman.etal}. In both aggregation-annihilation and
aggregation-fragmentation processes, rich kinetic behaviours arise due
to the competing reaction mechanisms.

In this paper we focus on fragmentation processes with mass-loss.
A number of authors have considered models combining fragmentation and 
{\it linear-annihilation} processes (i.e.\ a {\it sink}, where a species 
`evaporates' at a constant rate in time) 
(see, e.g., \cite{mcgrady.ziff88,edwards,machado} and references therein).
These models are relevant, for example, to studying oxidation and 
dissolution of solid porous media \cite{edwards,machado}.
However, to the best of our knowledge, the interplay between annihilation 
and fragmentation remains unexplored in the case of
{\it binary-annihilation} reactions. 
It is this subject with which we are concerned.
The kinetics of these phenomena are {\it non-linear}, and therefore more 
complex and, in principle, more difficult to analyse.
Fragmenting systems where fragments react at collision (maybe in the
presence of some agent) to produce an inert species, would yield
potential applications of these combined processes.
We investigate, at the mean-field level, a class of (non-linear) 
annihilation-fragmentation models, aimed at elucidating the
competition between these processes.
For all models considered we include, for comparison, an analysis of
the corresponding model with linear-annihilation (a sink).

In the simplest situation of a single species, the cluster-density decays 
as in the case of pure-annihilation, while the cluster-mass distribution 
approaches an exponential scaling distribution.
An exponential decay of the scaling distribution has also been found in
pure-fragmentation, aggregation-annihilation \cite{bennaim.krap95,krap93}
and reversible-aggregation \cite{family.etal}.
To allow for more interesting, and possibly more realistic
behaviour, we introduce additional species which work as catalysts for
annihilation and/or fragmentation; reactions between the catalysts are
also allowed for.
We find that the density-decay of the main-species can be surprisingly
unconventional, depending on the specific model and on the initial
densities of the catalysts. The decay may be universal but contain
logarithmic factors, or it may contain a power-law in time with a
non-universal exponent which depends on the reaction rates.
The interplay between annihilation and fragmentation and the existence
of a scaling regime, are determined by the asymptotic behaviour of the
mass-density and typical cluster-mass. These may either decay
indefinitely or tend to a constant value, depending on the relative
amounts of annihilation and fragmentation catalysts.

This paper is organized as follows.
In Sec.\ 2 we introduce and solve a model with single-species
fragmentation-annihilation where the fragmentation kernel has a rather
general form. The results for the corresponding model with linear
annihilation are stated for comparison.
In Sec.\ 3 we include catalysts in our model and solve it
for general time-dependences of the catalysts. Specific asymptotic
results are presented in the case where the fragmentation and
annihilation catalysts mutually annihilate.
The conclusions and further comments and suggestions are given in Sec.\ 4.

\section{Single-species annihilation-fragmentation}

\subsection{non-linear annihilation (model 1)}

The more elementary model combining the processes of pair-annihilation
and fragmentation is that involving a single `molecular' species.
This is described by the binary reaction scheme
\begin{equation}
  A_i + A_j \rightarrow inert \ , \ A_{i+j} \rightarrow A_i + A_j \ ,
\label{m1}
\end{equation}
where $A_i$ denotes a cluster consisting of $i$ monomers of species $A$.
The following calculations could be repeated for a discrete model,
but for analytical convenience we shall take the continuous limit,
which is standard practice \cite{ziff.mcgrady86,cheng.redner90}.

Let $a(x,t)$ be the density distribution of cluster-masses (or sizes)
$x$ at time $t$, and
\begin{equation}
  A_n(t)=\int^{\infty}_0 dx x^n a(x,t) \label{A.def}
\end{equation}
its moments.
In particular, the zeroth and first moments, $A(t)\equiv A_0(t)$ and
$A_1(t)$, denote the {\it total number-density} of clusters and the
{\it total mass-density}, respectively.
For simplicity, we consider linear fragmentation (driven by
a homogeneous external agent \cite{cheng.redner90}), neglect multiple
breakup, and take the rate that a cluster of mass $x$ breaks into
clusters of mass $x-y$ and $y$, $F(y,x-y)$, to be a constant (which we
set to be unity by a rescaling of time).
Hence, the total rate of fragmentation of any cluster,
$\int_0^x dyF(y,x-y)$, is equal to its mass, $x$.
We also envisage the simplest possible reaction, where two clusters
of any mass mutually annihilate upon encounter at a rate independent of
their masses.
Later, we shall see how to generalize our results for a kernel of the
form $F(y,x-y)=x^{\lambda-1}$, independent of the masses of the fragments.

Treating the annihilation term within the mean-field description,
which is valid in the reaction-controlled regime, the kinetic equation
for $a(x,t)$ reads
\begin{equation}
  \dot{a}(x) = -xa(x) + 2\int_x^{\infty}dy a(y) - J a(x) A \ , \label{a.eq}
\end{equation}
where $J/2$ is the relative rate of annihilation (equal rates of
fragmentation and annihilation correspond to $J=2$).
A set of non-linear equations for the moments of $a(x)$ follows immediately,
\begin{equation}
  \dot{A}_n= \frac{1-n}{1+n} A_{n+1} - JA_nA \ . \label{An.eq}
\end{equation}
A closed set of equations is found for the first two moments,
\begin{eqnarray}
  \dot{A}   & = & A_1 - J A^2 \ , \label{A.eq} \\
  \dot{A_1} & = &     - J A A_1 \ , \label{A1.eq}
\end{eqnarray}
whilst the higher moments obey an infinite hierarchy of equations which
are, apparently, not soluble.
The absence of a linear term in (\ref{A1.eq}) is a reflection of the
conservation of mass in the fragmentation process.
Adopting more general rates of fragmentation and annihilation,
say $x^{\lambda}$ and $(xy)^{\alpha}$, respectively, would yield the
following generalization of (\ref{An.eq}):
\begin{equation}
  \dot{A}_n= \frac{1-n}{1+n} A_{n+\lambda} - JA_{n+\alpha}A_{\alpha} \ .
\label{An.eq.gen}
\end{equation}
Clearly, no closed set of equations can be obtained for $A$ and $A_1$,
unless $\lambda=1$ and $\alpha=0$, which justifies our present choice.

To solve for $A(t)$ and $A_1(t)$, we proceed as follows. From 
(\ref{A.eq})-(\ref{A1.eq}), the {\it cluster average-mass},
$L(t) \equiv A_1(t)/A(t)$, obeys the equation
\begin{equation}
  \dot{L} = - L^2 \ , \label{L.eq}
\end{equation}
with solution
\begin{equation}
  L(t) = \frac{1}{\delta + t} \ \simeq \ \frac{1}{t} \ \ , \ \
  (t\to\infty) \ , \label{L.sol}
\end{equation}
where $\delta\equiv A_0/A_{10}=1/L(0)$ and $A_0\equiv A(0)$ and
$A_{10}\equiv A_1(0)$.
Using this result, Eqs.\ (\ref{A.eq})-(\ref{A1.eq}) can be
solved to give
\begin{eqnarray}
  A(t) & = & \frac{A_0 + A_{10}t}{1+J[A_0t+A_{10}t^2/2]}
  \ \simeq \ \frac{2}{Jt} \ , \label{A.sol} \\
  A_1(t) & = & \frac{A(t)}{\delta + t}
  \ \simeq \ \frac{2}{Jt^2} \ . \label{A1.sol}
\end{eqnarray}
The asymptotic behaviour in (\ref{A.sol})-(\ref{A1.sol}), and
hereafter, hold at large $t$.

To determine the mass distribution $a(x,t)$, we map the problem onto a
pure-fragmentation one using the auxiliary function
\begin{equation}
  b(x,t)\equiv a(x,t)\exp\left[J\int_0^tdt'A(t')\right] \ . \label{a.b}
\end{equation}
The exponential factor is known and from (\ref{A.sol}) we see
immediately that
\begin{equation}
\exp\left[J\int_0^tdt'A(t')\right]=1+J[A_0t+A_{10}t^2/2] \ . \label{exp}
\end{equation}
The function $b(x,t)$ obeys a standard fragmentation equation, i.e.\ it
is the same as (\ref{a.eq}) but without the non-linear annihilation term.
Its general solution is well known \cite{ziff.mcgrady85,saito}, and can
be written as
\begin{equation}
  b(x,t) = \left(b(x,0)+\int_x^{\infty}dy\,b(y,0)[2t+t^2(y-x)] \right)
  \:\exp(-tx) \ . \label{b.gen}
\end{equation}
We find it easier and more illustrative, however, to work from a
particular initial condition, and then to extract the asymptotic,
universal behaviour from the special solution obtained.
We adopt the exponential initial distribution
\begin{equation}
  a(x,0) = b(x,0) = \beta_0 \exp(-b_0 x) \ , \label{ax.0}
\end{equation}
where $b_0=\delta$ and $\beta_0=A_0\delta$. The corresponding solution
is obtained by substituting (\ref{ax.0}) in (\ref{b.gen}) or, equivalently,
attempting the ansatz $b(x,t) = \beta(t) \exp[-b(t)x]$ in the equation
for $b(x,t)$ \cite{redner}. This gives, after including the
time-dependent factor (\ref{exp}),
\begin{equation}
  a(x,t) \ = \ \frac{A_{10}(\delta+t)^2}{1+J[A_0t+A_{10}t^2/2]}
  \:\exp\left[-(\delta+t)x\right]
  \ = \ \frac{A(t)}{L(t)}\:\exp\left[-(\delta+t)x\right] \ . \label{ax.sol}
\end{equation}
As $t\to\infty$, $a(x,t)\simeq (2/J)\,\exp(-tx)$ becomes independent of
the initial condition. This universal asymptotic behaviour can also be
obtained by looking for a scaling solution of the form
\begin{equation}
  a(x,t) = L^{-w} \phi(\zeta) \ \ , \ \ \zeta = x/L \ , \label{ax.hyp}
\end{equation}
which we expect to hold for general initial conditions, in the regime
when $t$ is large and $x$ is small, with $xt$ an arbitrary constant.
 From (\ref{A.sol})-(\ref{A1.sol}) at large $t$, we find $w=0$.
Replacing $a$ by $\phi$ in Eq.\ (\ref{a.eq}) and taking the limit 
$t\to\infty$, yields a linear differential equation with solution
\begin{equation}
  a(x,t) = \phi(\zeta) = \frac{2}{J} \exp(-\zeta) \ . \label{ax.scal}
\end{equation}

One can calculate all the moments of $a(x,t)$ given by the special
solution (\ref{ax.sol}).
This yields for $A(t)$ and $A_1(t)$ exactly the same expressions as in
(\ref{A.sol})-(\ref{A1.sol}), which hold for general initial conditions
(One can check that this is also true for modified exponential initial
conditions, e.g.\ $x^p\exp[-b_0x]$.).
Since Eq.\ (\ref{An.eq}) for all moments is recursive, we conclude that
the special solution (\ref{ax.sol}) must have the same moments as the
general solution.
Hence, we did not need to solve for $A(t)$ in the first place;
rather, we could have solved self-consistently for $A(t)$ from this
special distribution.
The rationale is: that one may be able to determine a particular mass
distribution even when $A(t)$ is not known {\it a priori}, i.e.\ even
when there is not a closed set of equations as in the more general case
of Eq.\ (\ref{An.eq.gen}).

With this insight, we shall now solve the problem for a more general
fragmentation kernel, $F(y,x-y)=x^{\lambda-1}$, with $\lambda >0$.
There is no reason why the previous statement should not hold here,
i.e.\ we expect that for general $\lambda$ a particular exponential
solution will have the same moments as the general solution.
Eq.\ (\ref{a.eq}) now becomes
\begin{equation}
  \dot{a}(x) = -x^{\lambda}a(x) +
  2\int_x^{\infty}dyy^{\lambda-1} a(y) - J a(x) A \ . \label{a.eq.1}
\end{equation}
Once again we use the auxiliary function $b(x,t)$, given by
(\ref{a.b}), although now the exponential factor is not known.
This gives a pure-fragmentation equation for $b(x,t)$,
\begin{equation}
  \dot{b}(x) = -x^{\lambda}b(x) + 2\int_x^{\infty}dyy^{\lambda-1} b(y) \ ,
\label{b.eq}
\end{equation}
which has a known, but complicated, general solution \cite{ziff.mcgrady85}.
It has been proven that $\lambda >0$ is a necessary and sufficient condition
for $b(x,t)$ to have a scaling, universal behaviour
\cite{cheng.redner90,filippov}. We are allowed, therefore, to deduce
this behaviour from a special solution to (\ref{a.eq.1}).
A simple solution with exponential form (whose initial condition is an
obvious generalization of (\ref{ax.0})), is
\begin{equation}
  b(x,t) = \frac{A_{10}\lambda}{\Gamma(2/\lambda)}\,(b_0+t)^{2/\lambda}
           \exp[-(b_0+t) x^{\lambda}] \ , \label{bx}
\end{equation}
with moments
\begin{equation}
  B_n(t) = \frac{A_{10}\Lambda_{n+1,2}}{(b_0+t)^{(n-1)/\lambda}} \ .
\label{Bn}
\end{equation}
In (\ref{bx}) and (\ref{Bn}) appropriate normalization has been used.
We now have
\begin{equation}
  b_0^{1/\lambda} = \delta \Lambda_{2,1} \ \ , \ \
  \beta_0 = b_0^{2/\lambda} \frac{A_{10} \lambda}{\Gamma(2/\lambda)} \ ,
\label{b0}
\end{equation}
with the convenient notation
\begin{equation}
 \Lambda_{p,q} = \Lambda_{p,q}(\lambda) \equiv
 \frac{\Gamma(p/\lambda)}{\Gamma(q/\lambda)} \ . \label{Lambda}
\end{equation}

To find the explicit solution to the problem, we must determine the
density of clusters, $A(t)$, self-consistently.
 From (\ref{Bn}) and the transformation (\ref{a.b}) we obtain an
integral equation. This can be recast as the following differential equation
\begin{equation}
 \dot{A} = \frac{A}{\lambda(b_0+t)} - J A^2 \ , \label{A.eq.1}
\end{equation}
with solution
\begin{equation}
 A(t) \ = \ \frac{A_0 (1+t/b_0)^{1/\lambda}}
  {1+J A_0 b_0\frac{\lambda}{1+\lambda}
  \left[(1+t/b_0)^{(1+\lambda)/\lambda}-1\right]}
  \ \simeq \ \frac{1+\lambda}{\lambda}\frac{1}{Jt} \ \ , \ \ (t\to\infty) \ .
\label{A.sol.1}
\end{equation}
It is easy to see that $\exp(J\int_0^tdt'A)$ is simply the denominator
of $A$ in (\ref{A.sol.1}). From this, (\ref{a.b}) and (\ref{Bn}), we find
\begin{eqnarray}
 L(t) & = & \frac{\Lambda_{2,1}}{(b_0+t)^{1/\lambda}}
  \ \sim \ \frac{1}{t^{\lambda}} \ \ , \ \ (t\to\infty) \ , \nonumber \\
 A_n(t) & = & \Lambda_{n+1,1} \frac{A(t)}{(b_0+t)^{n/\lambda}}
  \ \sim \ \frac{\lambda+1}{\lambda} \frac{1}{J\,t^{1+n/\lambda}} \ ,
\label{sol.1} \\
 a(x,t) & = & \frac{\lambda\Lambda_{2,1}}{\Gamma(1/\lambda)}
  \:\frac{A(t)}{L(t)}\:\exp[-(b_0+t)x^{\lambda}]
  \ \sim \ \frac{\exp[-\zeta^{\lambda}]}{J\,t\,L(t)} \ , \nonumber
\end{eqnarray}
with $\zeta = \Lambda_{2,1}\, x/L(t)$.
All results reduce to the previous ones if we set $\lambda=1$.
Apart from numerical factors, the essential modifications in the asymptotics
can be alluded to the time-dependence of the average cluster-size,
$L\sim t^{-1/\lambda}$, and the dependence of $a(x,t)$ on the scaling
variable $\zeta$. We see that an increase in $\lambda$ slows
down the decay of $L$. This follows from the resulting decrease in the
rate of fragmentation of small fragments.

The mass distribution in (\ref{sol.1}), or (\ref{ax.scal}), is peaked
around $\zeta=0$, showing that the majority of the clusters have
vanishingly-small mass at late times due to fragmentation.
The factor $1/J$ indicates that if the annihilation rate increases the
mass and cluster densities should decrease.
The cluster density $A(t)$ is the only moment whose decay is
independent of $\lambda$. We recall that in pure-annihilation processes
$A \simeq 1/Jt$, while the cluster-mass distribution remains invariant
in time, apart from an overall time-dependent factor.
Our results indicate, therefore, that the fragmentation-annihilation
process at late-times consists essentially of the annihilation of
small-mass clusters.
This leads us to conjecture that in this case the upper critical
dimension is 2, as in the pure-annihilating case. That is clearly
true for a discrete system, where there is a minimum cluster-mass and
the fragmentation process ends before all particles annihilate.

\subsection{linear annihilation (model 2)}

Here, we briefly discuss model 2: a model with the same
fragmentation process as model 1 but with {\it linear} annihilation.
In this case there are no binary reactions and single clusters
annihilate at a constant rate in time.
We introduce this model for comparison with model 1 and because in Sec.\ 3
we shall make an interesting generalization of it.

The kinetic equation for the cluster-mass distribution is similar to
(\ref{a.eq.1}), but with the annihilation term replaced by the linear
term $-J a(x)$.
This yields a trivial modification of the pure-fragmentation model
(\ref{b.eq}) for general $\lambda$: the transformation (\ref{a.b})
is replaced by $b(x,t)=a(x,t)\,\exp[Jt]$, leading to a simple extra
exponential factor in the results.
Working once again with the special solution (\ref{bx}), which
corresponds to the initial condition
$a(x,0)=b(x,0)=\beta_0\,\exp[-b_0x^{\lambda}]$, we obtain
\begin{eqnarray}
 A(t) & = & \frac{A_{10}}{L(t)}\:\exp[-Jt] \ \sim \ t^{1/\lambda}\:\exp[-Jt]
 \ \ , \ \ (t\to\infty) \ , \nonumber \\
 A_n(t) & = & \Lambda_{n+1,1} \frac{A(t)}{(b_0+t)^{n/\lambda}}
  \ \sim \ \frac{\exp[-Jt]}{t^{(n-1)/\lambda}} \ ,
\label{sol.2} \\
 a(x,t) & = & \frac{\lambda\Lambda_{2,1}}{\Gamma(1/\lambda)}
  \:\frac{A(t)}{L(t)}\:\exp[-(b_0+t)x^{\lambda}]
  \ \sim \ \frac{\exp[-\zeta^{\lambda}]\:\exp[-Jt]}{L(t)^2} \ , \nonumber
\end{eqnarray}
where $L(t)$ is the same as in model 1 (Eq.\ (\ref{sol.1})).
Similar remarks about the dependence on $\lambda$ apply here.

\section{Annihilation-fragmentation with catalysts}

Model 1 has elucidated some of the features in the interplay
between the fragmentation and annihilation processes.
However, it appears to have a somewhat unrealistic drawback:
when two clusters of species A meet they react to produce a new,
inert species, rather than merging into a larger cluster.
Such a reaction would seem plausible, though, if another species, say
$C$, was involved. In a reaction of the form
$ A_x + C \rightarrow inert $, or $ A_x + A_y + C \rightarrow inert $,
the kinetic equations of $A$  and $C$ would be coupled together.
To be able to make progress, however, we envisage a situation where
the reactions are induced by the presence of $C$ but $C$ does not
take part in the reactions, i.e.\ $C$ is a catalyst.
In Sec.\ 2, it was implicitly assumed that annihilation was driven
by a catalyst of constant concentration. Here we shall allow the
concentration of $C$ to vary in time.
Catalysts play a key role in chemical, biological and ecological
systems (see, e.g., \cite{murray,prigogine}), so the study of this
class of models is potentially of great interest.

With these insights, let us consider the following two models where
the fragmentation of $A$ is controlled by a catalyst $B$,
\begin{equation}
  A_{x+y} + B \rightarrow A_x + A_y + B \ , \label{frag.B}
\end{equation}
and the annihilation of $A$ is induced by a catalyst $C$.
With a {\it non-linear annihilation} reaction ({\it model C1})
\begin{equation}
  A_x + A_y + C \rightarrow inert + C \ , \label{anni.1}
\end{equation}
while with a {\it linear annihilation} reaction ({\it model C2})
\begin{equation}
  A_x + C \rightarrow inert + C \ . \label{anni.2}
\end{equation}
In addition, we allow the catalysts to undergo the following
annihilation reaction, at rate R:
\begin{equation}
  B + C \stackrel{R}{\rightarrow} inert \ . \label{anni.BC}
\end{equation}
In the reaction-controlled regime $B(t)$ and $C(t)$, the concentrations
of species $B$ and $C$, have the following time dependence
(see, e.g., \cite{kang.redner}):
\begin{eqnarray}
B(t) \ = \ C(t) \ = \ \frac{B_0}{1+B_0Rt} \ \ , \ \ (B_0=C_0) \ , \nonumber \\
B(t) \ = \ \frac{\Delta B_0\exp(-\Delta Rt)}{C_0-B_0\exp(-\Delta Rt)}
\ \ , \ \ C(t) \ = \ \Delta + B(t) \ \ , \ \ (C_0>B_0) \ , \label{BC}
\end{eqnarray}
with $B_0\equiv B(0)$, $C_0\equiv C(0)$ and $\Delta\equiv |C_0-B_0|$.
Note that if $B(t)$ and $C(t)$ were constant in time, then, via a trivial
rescaling, models C1 and C2 would reduce to models 1 and 2 without catalysts.

\subsection{method of solution}

We shall solve the generalizations of models 1 and 2, with general values
of $\lambda$, following a similar procedure to that used in Sec.\ 2.
The essential difference now, is that a formal time variable has to
be defined, $\tau(t)$, which plays the role of time $t$ in the previous
models.

Model C1, for non-linear annihilation in the presence of
catalysts, is defined by the reaction schemes (\ref{frag.B}) and
(\ref{anni.1}). The kinetic equation for the cluster-mass
distribution, which generalizes (\ref{a.eq.1}), is
\begin{equation}
  \dot{a}(x) = \left[ -x^{\lambda}a(x) +
  2\int_x^{\infty}dy y^{\lambda-1} a(y) \right] B - J a(x) A C \ ,
\label{a.eq.c1}
\end{equation}
while the equation for the moments of $a(x,t)$ now reads (cf.\
(\ref{An.eq}) with $\alpha=0$)
\begin{equation}
  \dot{A}_n= \frac{1-n}{1+n}\,A_{n+\lambda}\,B - JA_{n}\,A\,C \ .
\label{An.eq.c1}
\end{equation}
For $\lambda=1$, a closed set of equations can be obtained once again,
 from which the general solution for $A(t)$ and $A_1(t)$ can be found.
For general $\lambda$, we map (\ref{a.eq.c1}) onto a pure-fragmentation
equation using the transformation (cf.\ (\ref{a.b}))
\begin{equation}
  b(x,t)\equiv a(x,t)\exp\left[J\int_0^tdt'A(t')\,C(t') \right] \ ,
\label{a.b.c1}
\end{equation}
and redefining time as
\begin{equation}
  \tau(t) = \int_0^tdt'\,B(t') \ . \label{tau}
\end{equation}
Clearly, one only expects a scaling regime to occur if
$\tau(t)\to\infty$ as $t\to\infty$.
Then $b(x,\tau)$ obeys Eq.\ (\ref{b.eq}). As usual, we work with
the special exponential solution (\ref{bx}) (now with $\tau$ replacing
$t$). Inserting this solution into (\ref{a.b.c1}) and differentiating
once, yields the self-consistent equation
(cf.\ (\ref{A.eq.1})-(\ref{A.sol.1}))
\begin{equation}
 \dot{A} = \frac{AB}{\lambda(b_0+\tau)} - J A^2 C \ , \label{A.eq.c1}
\end{equation}
with solution
\begin{equation}
 A(t) = \frac{A_{10}\Lambda_{1,2} (b_0+\tau)^{1/\lambda}}
{1+JA_{10}\Lambda_{1,2}\int_0^tdt'C(t')[b_0+\tau(t')]^{1/\lambda}} \ .
\label{A.sol.c1}
\end{equation}
As in Sec.\ 2, $\exp(J\int_0^tdt'AC)$ is simply the denominator of $A$
in (\ref{A.sol.c1}). Hence, from (\ref{bx}), (\ref{Bn}) and (\ref{a.b.c1})
we find
\begin{eqnarray}
 L(t) & = & \frac{\Lambda_{2,1}}{(b_0+\tau)^{1/\lambda}} \ , \nonumber \\
 A_n(t) & = & \Lambda_{n+1,1} \frac{A(t)}{(b_0+\tau)^{n/\lambda}} \ ,
\label{sol.c1} \\
 a(x,t) & = & \frac{\lambda\Lambda_{21}}{\Gamma(1/\lambda)}
  \:\frac{A(t)}{L(t)}\:\exp[-(b_0+\tau)x^{\lambda}] \ . \nonumber
\end{eqnarray}
The essential differences from (\ref{sol.1}) are
due to the presence of $\tau(t)$.

We now turn to model C2, for linear annihilation in the presence of
catalysts, defined by the reaction schemes (\ref{frag.B}) and
(\ref{anni.2}). The kinetic equation for $a(x,t)$ is similar to
(\ref{a.eq.c1}), but with the non-linear term replaced by $-Ja(x)C$. Defining
\begin{equation}
  b(x,t)\equiv a(x,t)\exp\left[J\int_0^tdt'C(t') \right] \ ,
\label{a.b.c2}
\end{equation}
and using the same formal time variable $\tau(t)$, $b(x,\tau)$
obeys Eq.\ (\ref{b.eq}) once again, and we work with the special
exponential solution as before.
The results, therefore, follow from (\ref{bx})-(\ref{Bn}) (with $\tau$
replacing $t$) multiplied by $\exp[-J\int_0^tdt'C]$. This gives
(cf.\ (\ref{sol.2}))
\begin{equation}
 A(t) = \frac{A_{10}}{L(t)}\:\exp[-J\int_0^tdt'C(t')] \ ,
\label{A.sol.c2}
\end{equation}
with $L(t)$, $A_n(t)$ and $a(x,t)$ still given by (\ref{sol.c1}).

\subsection{asymptotic results for $B+C\rightarrow inert$}

The results derived in Sec.\ 3.1 for models C1 and C2 hold for any
time dependence of the catalysts.
To obtain the asymptotic behaviours corresponding to catalysts $B$ and
$C$ mutually annihilating according to (\ref{anni.BC}), we insert the
mean-field expressions (\ref{BC}) into the previous results and look
for the dominant contributions in the limit $t\to\infty$.
There are three qualitatively different cases, depending on the relative
values of the initial densities of the catalysts. \\

\underline{{\bf $B_0=C_0$}}

With balanced amounts of the two catalysts, we have
\begin{equation}
  \tau \ = \ \ln(1+RB_0t)^{1/R} \ \simeq \ \ln(t^{1/R})
  \ \ , \ \ (t\to\infty) \label{tau.bc} \ .
\end{equation}
This gives,
\begin{eqnarray}
  L(t) & \simeq & \frac{\Lambda_{2,1}\,R^{1/\lambda}}
  {\left[\ln(t)\right]^{1/\lambda}} \ \ , \ \ (t\to\infty) \ , \nonumber \\
  A_n(t) & \sim & L(t)^n\, A(t) \ ,
\label{asym.bc} \\
  a(x,t) & \sim & \frac{A(t)}{L(t)}\:\exp[-\zeta^{\lambda}]
  \ \ , \ \ \zeta = \Lambda_{2,1}x/L(t) \ , \nonumber
\end{eqnarray}
for models C1 and C2, and
\begin{eqnarray}
  A(t) & \simeq & \frac{1+\lambda}{\lambda}\:\frac{R}{J\ln(t)}
 \ \sim \ \frac{1}{\ln(t^{J/R})} \ \ , \ \ \ \ \ \ {\rm model \ C1} \nonumber \\
       & \simeq & \frac{A_{10}}{L(t)}\:\frac{1}{[B_0Rt]^{J/R}}
  \ \sim \ \frac{[\ln(t)]^{1/\lambda}}{t^{J/R}} \ \ , \ \ {\rm model \ C2} \ .
\label{A.asym.bc}
\end{eqnarray}
As in models 1 and 2 without catalysts,
the late-time regime is dominated by the annihilation of small-mass
clusters and there is scaling.
The cluster-number density, however, decays logarithmically in model C1
and with a non-universal power-law in model C2: the
non-universal exponent depends on the annihilation rates $J$ and $R$ of
species $A$ and of the catalysts.
We conclude that, although different time-dependences emerge, the balance
between the annihilation and fragmentation processes in this symmetric case
is analogous to that occurring in the absence of catalysts. \\

\underline{{\bf $B_0>C_0$}}

If there is a majority of the fragmentation catalyst the auxiliary time
$\tau$ behaves essentially like $t$,
\begin{equation}
  \tau \ = \ \Delta t + \ln\{(B_0-C_0\exp[-\Delta Rt])/\Delta\}^{1/R}
  \ \simeq \ \Delta t  \ \ , \ \ (t\to\infty) \ .
\label{tau.b}
\end{equation}
The key point to notice when evaluating the integral in the denominator
of (\ref{A.sol.c1}) and the exponential term in (\ref{A.sol.c2}), is
that $\int_0^tdt'C(t')\to\ln(B_0/\Delta)^{1/R}$ as $t\to\infty$.
This yields, for models C1 and C2,
\begin{eqnarray}
  L(t) & \simeq &
   \frac{\Lambda_{2,1}}{(\Delta t)^{1/\lambda}}
   \ \ , \ \ (t\to\infty) \ , \nonumber \\
  A(t) & \simeq &
  A_{10}F(\lambda)\Lambda_{2,1}\:(\Delta t)^{1/\lambda} \ , \label{asym.b} \\
  A_1(t) & \to & A_{10}F(\lambda) \ , \nonumber \\
  a(x,t) & \sim &  \frac{1}{L(t)^2}\:\exp[-\zeta^{\lambda}]
  \ \ , \ \ \zeta = \Lambda_{2,1}x/L(t) \ , \nonumber
\end{eqnarray}
where
\begin{eqnarray}
  F(\lambda) & = & \frac{1}{1+JA_{10}\Lambda_{2,1}I_{\infty}}
  \ \ , \ \ {\rm model \ C1} \nonumber \\
  & = & \left(\frac{\Delta}{B_0}\right)^{J/R}
  \ \ , \ \ \ \ \ \ \ \ {\rm model \ C2}
\label{F.b} \\
  I_{\infty} & \equiv & \int_0^{\ln(B_0/\Delta)^{1/R}}dx
  \left[b_0+x+\ln\{C_0/(B_0-\Delta\exp[Rx])\}^{1/R}\right]^{1/\lambda} \ .
\label{I}
\end{eqnarray}
In this case the fragmentation process is much faster than the
annihilation one, which is suppressed as the mass-density
tends to a fraction $F(\lambda)$ of its initial value $A_{10}$.
Once again there a scaling regime.
In conclusion, the asymptotic behaviours of models C1 and C2 are similar
to that of a pure-fragmentation process with rescaled initial
mass-density $A_{10}F(\lambda)$ and rescaled time $\Delta t$, where $\Delta$
is the residue of catalyst $B$. Hence, the smaller the value of $\Delta$
the longer the time needed to reach this regime.
The value of $F(\lambda)$ is determined not only by the kinetic parameters
$J$, $R$ and $\lambda$, but also by the initial densities $B_0$ and $C_0$. \\

\underline{{\bf $C_0>B_0$}}

With a majority of the annihilation catalyst $\tau$ tends to a finite value,
\begin{equation}
  \tau \ = \ \ln\{(C_0-B_0\exp[-\Delta Rt])/\Delta\}^{1/R} \ \to \
  \tau_{\infty} \ = \ln(C_0/\Delta)^{1/R} \ \ , \ \ (t\to\infty)
\label{tau.c} \ .
\end{equation}
Noticing that the integral in the denominator of (\ref{A.sol.c1}),
$\int_0^tdt'C(t')[b_0+\tau(t')]^{1/\lambda}$, is dominated the large-$t$
contribution, we obtain
\begin{eqnarray}
  L(t) & \to & L_{\infty} \ = \ \Lambda_{2,1}/(b_0+\tau_{\infty})^{1/\lambda}
  \ \ , \ \ (t\to\infty) \ , \nonumber \\
  A_n(t) & \sim & A(t) \ , \label{asym.c} \\
  a(x,t) & \sim &  A(t)\:\exp[-(b_0+\tau_{\infty})x^{\lambda}] \ , \nonumber
\end{eqnarray}
for models C1 and C2, and (cf.\ (\ref{A.sol.1}) and (\ref{sol.2}))
\begin{eqnarray}
  A(t) & \simeq & \frac{1}{J\Delta t}
  \ \ , \ \ \ \ \ \ {\rm model \ C1} \ , \nonumber \\
  & \simeq & \frac{A_{10}}{L_{\infty}}\left(\frac{\Delta}{C_0}\right)^{J/R}
  \:\exp[-J\Delta t] \ \ , \ \ {\rm model \ C2}
\label{A.asym.c} \ .
\end{eqnarray}
Here, the annihilation process dominates over fragmentation.
The latter is suppressed as the average cluster-mass tends to a non-zero
value $L_{\infty}$ and the mass distribution becomes stationary (up to an
overall time-dependent factor) and identical to its initial form $a(x,0)$.
This form of $a(x,t)$ follows from all fragments being equally-likely
to annihilate, and the prefactor $A(t)\to 0$ indicates the eventual
loss of all mass. This situation differs from $B_0=C_0$, where
annihilation occurs between increasingly smaller clusters.
As expected, there is no scaling regime since $\tau$ has a finite limit.
In addition, the behaviour of $a(x,t)$ may not be universal in this case.
In conclusion, the asymptotic behaviour of models C1 and C2 is similar
to that of a pure-annihilation process with $L=L_{\infty}$ and rescaled
time $\Delta t$: the process being a binary reaction for model C1,
and linear with initial mass $A_{10}(\Delta/C_0)^{J/R}$ for model C2.
The asymptotic average-mass, $L_{\infty}$, is determined by the
kinetic parameters $R$ and $\lambda$, and by the initial values $A_0/A_{10}$,
$B_0$ and $C_0$.

It should be stressed that the validity of the continuous-mass description
used throughout is subject to restrictions in the long-time regime.
In the cases $B_0=C_0$ and $C_0>B_0$, where the cluster-number $A(t)\to
0$ and the system eventually reaches an empty state, the description holds
while the system still contains a sufficiently large number of particles.
This condition ensures the smoothness of the cluster-mass distribution.
In the cases $B_0=C_0$ and $B_0>C_0$, where the average cluster-mass
$L(t)\to 0$ and increasingly-smaller fragments are produced, the usual
restrictions to a pure-fragmentation process apply \cite{ziff.mcgrady86}.
The essential difference within a discrete description lays in the
behaviour of the density of the smaller fragments, or monomers.

\section{Conclusions}

We have solved a class of fragmentation models with non-linear
mass-loss, at the mean-field level, and elucidated the competition
between fragmentation and annihilation processes.
Simultaneously, we have analysed the corresponding models with
linear mass-loss.
Although the solutions for the cluster-mass distribution hold for
exponential initial conditions, we expect their moments to be
identical to those of the corresponding general solutions.

Models 1 and 2 (without catalysts) are characterized by the
annihilation of small-mass clusters at late times. We believe that in
this case the upper critical dimension is 2 (but see \cite{family.etal}).

Models C1 and C2, with fragmentation and annihilation catalysts,
have been solved for a general time-dependence of the catalyst
densities. The linear annihilation model, in particular, shows much
more interesting and less trivial behaviours in this case.
When $B$ and $C$ undergo mutual annihilation, we have found that
different interplays arise between the two intervening processes
depending on the initial densities $B_0$ and $C_0$. As a result,
unconventional and/or non-universal asymptotic behaviours may occur
($B_0=C_0$), or the scaling regime may be absent ($C_0>B_0$).
In these models, the upper critical dimension is likely to be 4,
as a consequence of the spatial structure induced by the catalytic
reactants at late times:
in $B$-rich regions fragmentation (annihilation) will be frequent (absent)
producing a high concentration of $A$, whilst in $C$-rich regions the
roles of these processes will be reversed producing voids of $A$.
With simultaneous fragmentation and annihilation it seems difficult,
however, to develop more quantitative arguments determining $d_c$.
In any case, new, unusual behaviours should also appear at {\it lower
dimensions} in the presence of catalysts.
Even here, the mean-field description may prove adequate up to
moderate times, before significant spatial fluctuations set in.

Instead of the mean-field expressions (\ref{BC}) for the catalyst densities,
one may use the corresponding ones in the diffusion-controlled regime
($d<4$), which can be deduced, e.g., from \cite{kang.redner}.
In addition, other reactions, or time-dependences of the catalysts may be
investigated. One may even replace one of the catalysts by an {\it inhibitor}.
For these purposes one simply needs to use the results of Sec.\ 3.1.
In practice, it may be that both linear and non-linear mass-loss
processes operate. In such cases, models 1(C) and 2(C) would merge to
yield a model with linear and non-linear annihilation terms. It should
be possible to solve such a model using the methods of Secs.\ 2 and 3.
It may also be that one has a source rather than a sink term, in which
case the previous terms will have opposite signs leading to a
competition between linear and non-linear contributions.

Further directions of work are possible, in addition to those just mentioned.
One may look for mass-dependent reaction-rates, the generic form for
the annihilation terms then being: $ - Ja(x)\int_0^{\infty}dyf(x,y)a(y)$
for model 1, and $-Ja(x)f(x)$ for model 2.
(In Sec.\ 2.1 we mentioned the case $f(x,y)=(xy)^{\alpha}$.)
Another research avenue would be to investigate two species, say $A$ and
$B$, undergoing fragmentation and mutual annihilation.
When $A_0=B_0$ this model is identical to the single-species model of
Sec.\ 2.1.
With two species, a `charge' conservation-law may also be present, e.g.,
$A_x+B_y\to A_{x-y}$ for $x>y$ ($B_{y-x}$ for $y>x$) \cite{krap93}.
We have found \cite{unpub} that in the absence of such a constraint both
cluster densities $A(t)$ and $B(t)$ decay as $2/(Jt)$ ($\lambda=1$),
independently of the initial masses $A_{10}$ and $B_{10}$,
i.e.\ there is no mass-difference residue.

We expect the classes of models studied here, and their further
developments, to have applications in various subjects, such as physics,
chemistry, biology and, to some extent, population dynamics (see, e.g.,
\cite{murray,prigogine}).
Model 1 (without catalysts) is suitable to describe fragmenting systems
with non-linear, reaction mass-loss.
Models C1 and C2, on the other hand, have the potential to describe
a broad range of phenomena, with linear or non-linear mass-loss and
with arbitrary time-dependences of the catalysts.
We describe one particularly important generalization to these models.
In many processes of industry, heterogeneous chemistry, or biochemistry,
such as food processing or food digestion (where the catalysts are
enzymes), the efficiency of the reactions may depend on the available
surface-area of the reacting fragments (see, e.g., \cite{fractals}).
It would be of great interest, therefore, to develop fragmentation
models whose clusters are characterized not only by their {\it volume}
(or mass), but also by their {\it surface-area}. From these, one could
then build up fragmentation-annihilation models with surface-area
dependent annihilation rates.
An obvious difficulty in constructing such models is the dependence
of the surface-area on the shape of the fragments.
First steps towards this goal has been taken in very recent work on
multivariable fragmentation \cite{multi}.

It is interesting to note that there is some resemblance between
fragmentation-annihilation processes and branching-annihilating random
walks \cite{branching}. In the latter, particles diffuse and annihilate
upon encounter and each particle gives birth to $n$ offspring, at
prescribed rates. A difference, though, is that the particles do not have
a mass distribution. Depending on the branching and annihilating rates,
these systems may evolve towards an empty state or an active steady-state.
The existence of phase transitions in the long-time behaviour has
been investigated for different values of $n$ \cite{branching}.
Depending on the densities of catalysts (which control the rates of
the two processes involved), we have found that models C1 and C2 can
also exhibit quite opposite asymptotic behaviours: $A\to 0$ for
$C_0\geq B_0$, and $A\to\infty$ for $B_0>C_0$.
Although we have not seen evidence for states characterized by steady
values of the cluster-number density $A(t)$, steady-state values were
obtained for the cluster-mass density $A_1(t)$ ($B_0>C_0$) and for the
average cluster-mass $L(t)$ ($C_0>B_0$).

In conclusion, we hope that this paper will stimulate future work on
fragmentation-annihilation processes, and believe that fruitful
applications of the present, or related models (especially those with
catalysts), can be found in various fields.

\section*{Acknowledgments}

We thank Gerald Lobley for having drawn our attention to the important 
role of the surface-area of fragments in the efficiency of protein 
metabolism and other nutrition processes.
We are indebted to Adil Hasan for a careful reading of the manuscript and
valuable remarks.

%\newpage
\singlespacing

\end{document}